\documentstyle[12pt]{article}
\newcommand{\beq}{\begin{equation}}
\newcommand{\eeq}{\end{equation}}
\newcommand{\eqna}{\begin{eqnarray}}
\newcommand{\eqne}{\end{eqnarray}}
\newcommand{\dia}{\begin{displaymath}}
\newcommand{\die}{\end{displaymath}}
\newcommand{\eqnaa}{\begin{eqnarray*}}
\newcommand{\eqnae}{\end{eqnarray*}}
\def\s{\Sigma}
\def\so{\Sigma_0}
\def\m{\mu}
\def\cit#1{$^{\cite{#1}}$}

\def\bk{\par\noindent}
\def\bk{\par\noindent}

\def\a{$a_0$}
\def\af{a_0}
\def\gn{g_N}
\begin{document}
\sloppy
\begin{center}	
{\tenrm\bf Second workshop on
 "The dark side of the Universe: experimental efforts and theoretical
framework", Rome, November 1995}\\
\vskip 0.5truein
{\tenrm\bf LOW-SURFACE-DENSITY GALAXIES AND THE MODIFIED DYNAMICS}\\
\vskip 0.5truein
{\tenrm\bf MORDEHAI MILGROM} \\
\bk
{\tenrm\it  Department of condensed-matter physics,
Weizmann Institute, Rehovot 76100, Israel}     \\
\vskip 1truecm
\baselineskip 12pt
{\tenrm\bf ABSTRACT}\\
\end{center}
{\tenrm Very-low-surface-density galactic systems have very low mean
 accelerations. They thus provide quintessential tests of the Modified
Dynamics (MOND), which predicts an increasing mass discrepancy with
decreasing acceleration.
We describe succinctly the results pertinent to several classes of such objects:
Low-luminosity (dwarf) spirals, irregular dwarf spirals,
 normal-lumonosity-but-low-surface-density
spirals, and dwarf-spheroidal satellites of the Milky Way.
\vskip 1truecm
\baselineskip 14pt

{\noindent\bf 1. The modified dynamics}
\vskip 0.5truecm
\par
As some of you may know, I have been advocating, with others,
 that there is
not much dark matter in galactic systems. The mass discrepancy
observed in galaxies is then due to a breakdown of Newtonian dynamics,
 which is used to determine the gravitational masses. The specific
alternative proposed\cit{mil83a,mil83b}, called MOND,
assumes that Newtonian dynamics (law of inertia and/or gravity)
break down when the acceleration of a test particle in a system
 is much smaller than some
borderline acceleration \a :
The Newtonian acceleration $\gn=MG/r^2$ that an attracting
mass $M$ produces on a test particle, a distance $r$ away from it,
is assumed to be valid only in the limit $\gn\gg\af$. In the opposite
limit, $\gn\ll\af$, the test-particle acceleration, $g$, is given
by $g^2/\af\approx\gn=MG/r^2$.
This basic idea may be interpreted as either a modification of Newtonian
gravity, or a modification of the law of inertia\cit{mil83a}, and can be
incorporated into Lagrangian theories in both the former\cit{bm84},
and the latter\cit{mil94a} interpretation.
There have been several attempts to develop relativistic extensions
for the modified-gravity approach( see e.g. refs.\cite{bek, bm84, san}), but
 none of these is without problems.
\par
The salient ramifications of MOND are captured by the
simplistic formulation that relate the acceleration $\vec g$ to
the acceleration, $\vec \gn$, calculated with Newtonian gravity,
by
\beq \m(g/\af)\vec g=\vec \gn,     \label {   } \eeq
where $g\equiv \vert\vec g\vert$, and $\m(x)$ is some extrapolating
function whose limiting behaviour at the two extreme values of its
argument are given by: $\m(x\gg 1) \approx 1$, to recover Newtonian
dynamics in this limit, and $\m(x\ll 1)\approx x$. Otherwise, $\m$
remains unspecified. The implications for galaxy dynamics
do not depend critically on the exact form of $\m$, as long as it
is assumed to be increasing. Accelerations in galactic systems
are never much larger than \a\ (see below) so $\m(x)$ has to be known
only up to $x$ of a few, in this context. In contrast, aspects such as
solar-system tests of the theory, which probe the region $g\gg\af$,
depend critically on just how fast $\m(x)$ approaches 1 at large $x$.
 For instance, the two choices $\m(x)=1-e^{-x}$, and $\m(x)=x/(1+x)$
make very similar predictions for galactic dynamics, but very
different ones for, say, the perihelion shift of planetary orbits:
The former predicts a totally negligible effect, while the latter
produces an effect that is already in conflict with the
 measurements\cit{mil83a}.
\par
The main predictions of MOND regarding galaxies are\cit{mil83b}:
\bk
1. The orbital velocity on a circular orbit far from a finite mass
is independent of the orbital radius. This leads to
 asymptotically flat rotation curves of disk galaxies.    \bk
2. The asymptotic velocity $V_{\infty}$, depends only on the total mass,
$M$, of the system (galaxy): $V_{\infty}^4=MG\af$. This gives
the Tully-Fisher relation. \bk
3. The mean velocity dispersion, $\sigma$, of a self gravitating system
 supported
 by random motions is strongly correlated with the total mass:
$\sigma^4\sim MG\af$. This leads to the Faber-Jackson relation for elliptical
galaxies. \bk
4. Thin galactic disks are more stable when their mean surface density,
 $\s$, satisfies
$\s\ll\so\equiv \af G^{-1}$
 (i.e. their mean acceleration is much smaller than \a).
 This explains the marked paucity of
galaxies with surface density above some cutoff value, known as the
 Freeman law \bk
5. Isothermal spheres do not exist that have a mean surface density
much exceeding $\so$. This accounts for the observed analouge of the
 Freeman law for elliptical galaxies, known as the Fish law. \bk
6. The rotation curve calculated for a galaxy using MOND, and
 assuming the presence of only the visible matter, should agree with
the observed rotation curves. This most detailed prediction was
tested repeatedly (see e.g. \cite{bbs91}).

\par
The constant \a\ appears in predictions 2-6 above, and can thus be
 determined
 (in several independent ways) by compaing the predictions with the
data. All these methods yield
$\af\sim (1-2)\times 10^{-8}~cm~sec^{-2}$.
\par
It may be most significant\cit{mil83a}  that \a\ turns out to be of the
 same order as $cH_0$ ($H_0$ being the Hubble constant).
 This may betoken some connection of MOND
with cosmology, in the spirit of Mach's principle (see more details in
\cite{mil94a}).

The mass discrepancy in clusters of galaxies (at radii of a few Mpc) is well
 accounted for by MOND (e.g. ref. \cite{san94}), and so is the discrepancy in
	small galaxy groups.

 The only place where MOND fails systematically to explain away
dynamical dark matter is in the cores (up to a few hundred kpc)
of rich, x-ray galaxy clusters (e.g. ref.\cite{gdl92}).
 Even with MOND,
most of the mass required in these cores by x-ray gas hydrostatics,
and by lensing, must be genuine, yet-undetected dark matter. Cooling flows are
 known to carry cool gas into these cores whose fate is not clear, and which	
have so far escaped detection.	
While the estimated, present-day mass deposition rates are too small
to account for the total dark mass required in the cores, they could have been
much larger in the past.

If, in fact, the mass discrepancy bespeaks the presence of DM and not of
new physiscs, then MOND is, in the least, a very
economical description of the mass distribution,
and tells us that the amount and distribution of DM in galaxies
uncannyly follows a very strict rule involving only one parameter
(\a), and is fully determined by the distribution of
visible matter. This is quite hard to believe.
\vskip 1truecm
{\noindent\bf 2. Low-surface-density galaxies}
\vskip 0.5truecm
\par
MOND was introduced to account for the behaviour of the
rotation curves of "normal" disc galaxies at large radii, where the
accelerations become very small. The
asympotics of rotation curves actually determine the essential
phenomenology of MOND:
The linear form of $\mu(x)$ for small $x$ is dictated by the asymptotic
flatness of rotation curves. The value of \a\ is fixed by the intercept of the
Tully-Fisher relation.
However, small accelerations are found in the realm of the galaxies not
only in the outskirts of galaxies. There are systems in which the
 accelerations are very small everywhere from the centre out. These are
the low-surface-density (LSD) galaxies, which are particularly crucial in
 testing the modified dynamics. This has to do with the fact that the
 mean surface density of a galaxy, $\s=M/\pi R^2$, is a direct measure of
its mean (Newtonian) acceleration $GM/R^2$.
Defining thus $\so\equiv G^{-1}\af$, we see that systems with
a mean surface density $\s\ll\so$ are deep in the MOND regime.
In predicting their behaviour no leeway is left in adjusting the theory.
Such system afford particularly sharp tests of MOND because
a. The expected mass discrepancy is large.
b. The dynamics is practically independent of the assumed form of
$\m(x)$ because in the relevant, $x\ll 1$, regime we have $\m\approx x$.
c. The {\it shape} of the rotation curves
predicted by MOND for an LSD galaxy
is independent of various galaxy parameters that are
 not always known with good accuracy: The
distance to the galaxy, its stellar $M/L$, its inclination, as well
as the value of \a\ all enter together only in the normalization of
the predicted curve.
d. Many LSD galaxies are dominated by gas mass (relative to stellar mass)			
and hence their analysis depends rather weakly on the assumed stelalr $M/L$ 	
values, yielding almost parameter-free MOND predictions.

\par
Various galaxy types fall in the class of LSD galaxies:
1. dwarf spirals
2. dwarf irregulars (described e.g. in \cite{lsy93}). One obtains
3. normal-, or high-luminosity spirals with low surface brightness
 (see e.g. ref.\cite{vdh93}).
4. The dwarf spheroidal satellites of the Milky Way with stellar $\s$
down to a few percent of $\so$  (see \cite{mat94} for a review).
\par
\vskip 1truecm
{\noindent\bf 3. MOND predictions and observations of LSD galaxies}
\vskip 0.5truecm
{\noindent\it 3.1 LSD, Dwarf Spirals}
\par
There are now quite a few dwarf spirals	 for which 			
the roataion curves, as well as the (stellar and gasious) mass distributions
have been measured. They afford very acute tests of MOND for the reasons
 explained above. They were predicted\cit{mil83b} to show a large mass
 discrepancy right from the center of the galaxy, long
 before any of the above data was available. Rotation-curve analysis of these
	show very good agreement with the RCs predicted by MOND. Some examples of
	these are given e.g. in ref.\cite{bbs91}.

\vskip 0.5truecm
{\noindent\it 3.2 Dwarf-Irregular Spirals}
\par
Milgrom\cit{mil94b} has analyzed the data of ref. \cite{lsy93} using
a generalized MOND virial relation that relates the total mass to the
rms velocity dispersion in LSD systems.
He found that the masses predicted by MOND from the observed dispersions agree
with the observed masses (gas plus stars with reasonable $M/L$  values of	
 order one solar unit). In contrast, the Newtonian $M/L$ values found in
 ref.\cite{lsy93} range between 7 and 26.

\vskip 0.5truecm
{\noindent\it 3.3 LSD, Normal-Luminosity Spirals}
\par
 McGaugh et al. (ref. \cite{mbh95}) have analyzed some twenty-five galaxies
 spanning a large range of surface densities, most of which
 having low surface densities in the sense we discuss here. They find a
strong correlation between the Newtonial $M/L$ value and their mean surface
 density, in just the way predicted by MOND: The-lower-surface-density galaxies
 have higher
 Newtonian $M/L$ values (up to a few tens solar units). Their MOND analysis of
the same sample gives a mean MOND $M/L$ value of order unity across the full
	surface-density range. 	

\vskip 0.5truecm
{\noindent\it 3.4 Dwarf-Spheroidal Satellites of the Milky Way}
\par
As was predicted by MOND\cit{mil83b} these are now known to
 evince large mass discrepancies when analyzed by
 Newtonian dynamics (see e.g. ref \cite{mat94}, and references therein).
In a recent MOND analysis, using updated velocity despersions for some
of the dwarfs, it was found\cit{mil95} that the dynamics is explained with
$M/L$ values typical of globular-cluster stellar populations, i.e.
with no need for dark matter. The results are summarized in Table 1, together
 with the estimated Newtonain $M/L$ values.
\begin{table}	
\begin{tabular}{|c|ccccccc|} \hline\hline
Dwarf galaxy & Sculptor & Sextans & Carina & Draco & LeoII & U Minor & Fornax\\
\hline\hline
Newtonian $ M/L $ & $\sim$ 12 & $\sim$ 18 & 16--62 & 50--120 & 7--15 &35--100 &
 5--26  \\
\hline
MOND $ M/L$ & 0.7--2 &0.7--3 & 1.5--7 & 3--6 & 0.7--4 & 2.5--6.5 & 0.1--1.4  \\
\hline
\end{tabular}
\caption{The estimated Newtonian and MOND $M/L$ ranges for the seven dwarf
 spheroidals with measured velocity dispersions}	
\end{table}

{\bf Acknowledgement}
I thank Stacy McGaugh for permission to quote from the results of
ref. \cite{mbh95} before their publication.


\begin{thebibliography}{999}
\bibitem{bbs91}
K.G. Begeman, A.H. Broeils, and R.H. Sanders 1991, MNRAS, 249, 523.
\bibitem{bek}J. Bekenstein 1992, in Proc. 6th Marcel Grossman Meeting on GR,
eds. H.Sato \& T. Nakamura (Singapore: World Scientific), 905.
\bibitem{bm84} J. Bekenstein, and M. Milgrom 1984, ApJ, 286, 7.
\bibitem{gdl92} D. Gerbal, F. Durret, M. Lachi\'ez-Ray, and G. Lima-Neto 1992,
A\& A, 262, 395.

\bibitem{lsy93}K.Y. Lo, W.L.W. Sargent, and K. Young 1993, AJ, 106, 507.
\bibitem{mat94}M. Mateo 1994, Proceedings of the ESO/OHP workshop
 ``Dwarf Galaxies'', eds. G. Meylan and P. Prugniel.
\bibitem{mbh95} S.S. McGaugh, W.J.G. de Blok, J.M. van der Hulst, and
M.A. Zwann preprint (1995).
\bibitem{mil83a} M. Milgrom, 1983a, ApJ, 270, 365.
\bibitem{mil83b} M. Milgrom, 1983b, ApJ, 270, 371.

\bibitem{mil94a} M. Milgrom 1994a, Annals of Physics, 229, 384.
\bibitem{mil94b} M. Milgrom 1994b, ApJ, 429, 540.
\bibitem{mil95} M. Milgrom 1995, ApJ, 455, 439.
\bibitem{san94}R.H. Sanders,  1994, A\& A, 284, L31.
\bibitem{san}R.H. Sanders,  1990, A\& A Rev., 2, 1.
\bibitem{vdh93}J.M. van der Hulst, E.D. Skillman, T.R. Smith, G.D. Bothun,
S.S. McGaugh, and W.J.G. de Blok 1993, AJ, 106, 548.


\end{thebibliography}
\end{document}